\documentclass[usenatbib]{jaa}

\usepackage{graphicx}
\usepackage{xcolor}
\usepackage{url}
\usepackage{aasmacros}

\begin{document}\sloppy

\title{Reconstruction of Dynamical Dark Energy Potentials:
  Quintessence, Tachyon and interacting models.} 

\author{Manvendra Pratap Rajvanshi\textsuperscript{1} and
  J.S.Bagla\textsuperscript{1}*}  
\affilOne{\textsuperscript{1}Indian Institute of Science Education and
  Research Mohali, \\ 
Knowledge City, Sector 81, Sahibzada Ajit Singh Nagar, Punjab 140306, 
India}

\twocolumn[{

\maketitle

\corres{jasjeet@iisermohali.ac.in}



\begin{abstract}
Dynamical models for dark energy are an alternative to the
cosmological constant.
It is important to investigate properties of perturbations in these
models and go beyond the smooth FRLW cosmology.
This allows us to distinguish different dark energy models with the
same expansion history.
For this, one often needs the potential for a particular
expansion history. 
We study how such potentials can be reconstructed obtaining closed
formulae for potential or reducing the problem to quadrature.  
We consider three classes of models here: tachyons, quintessence and
interacting dark energy.
We present results for constant $w$ and the CPL parameterization.
The method given here can be generalized to any arbitrary form of $w(z)$.  
\end{abstract}


}]

\year{yyyy} 
\pgrange{num--num}

\volnum{000}
\pgrange{1--}
\setcounter{page}{1}

\section{Introduction}

Observations (Riess {\em et al.} 1998; Perlmutter {\em et al.} 1999) have
indicated that the Universe is expanding at an increasing rate. 
This has led to Dark Energy
(Efstathiou {\em et al.} 1990; Ostriker \& Steinhardt 1995; Bagla {\em et al.} 1996; Amendola \& Tsujikawa 2010),
the component with unusual properties that causes the accelerated
expansion of the Universe.  
Besides the simple and successful Cosmological Constant model
($\Lambda$), there are a number of competing theories
(Amendola \& Tsujikawa 2010; Durrer 2011; Bamba {\em et al.} 2012) that are consistent with
observations.   
In these theories, dark energy is dynamical in that its
properties are a function of space and time.
In order to study the theoretical and observational implications for
these theories, we have to solve the equations describing the dark
energy.
Analysis of some observations only requires the variation of scale
factor with time, however other observations can have a dependence on
spatial variations in dark energy and thus details of the model
become relevant.

A number of models have been proposed for dark energy, e.g., tachyon 
dark energy (Padmanabhan 2002; Bagla {\em et al.} 2003) and
quintessence (Tsujikawa 2013; Caldwell {\em et al.} 1998). 
In both of these a scalar field and its gradients give rise to dark
energy densities, but the forms of the Lagrangian densities for these
are very different.

It is well known that if two models have the same evolution of the
scale factor, tests relying only on distance measurements cannot
distinguish between such models.
Therefore it is important to study growth of perturbations in matter
for different models of dark energy with the same evolution of the
scale factor.
This opens up comparison based on CMB anisotropies
(Jassal 2012; Wang {\em et al.} 2010; Mifsud \& De Bruck 2017), weak lensing
(Bernardis {\em et al.} 2011; Pettorino \& Baccigalupi 2008; Yang {\em et al.} 2016), and growth of
perturbations (Rajvanshi \& Bagla 2018).  
In this context, it is useful to have a formalism for constructing
potentials for different models of dark energy that lead to the same
expansion history. 
In this article we compute the corresponding potentials in quintessence
and tachyon models which can give same background evolution.
We reconstruct potential $V(\phi)$ assuming a particular equation of
state $w(z)$.
We give analytical expressions wherever possible, in other cases we
reduce the problem to quadrature for numerical reconstruction of
$V(\phi)$. 

There has been a lot of interest in recovering dark energy
potential from the observed expansion history
(Wu \& Yu 2007; Sahni \& Starobinsky 2006; Saini {\em et al.} 1999). 
For example Huterer and Turner (Huterer \& Turner 1999), provide an
early work on constructing potential from simulated data and inspired
further research.
Li et.al (Li {\em et al.} 2007) construct potential by
approximating luminosity distances and also do a comparison for
reconstruction using parameterization of equation of state $w(z)$.
A number of other attempts for reconstruction using a parametric or a
non-parametric approach have been made. 
For example, see
(Gerke \& Efstathiou 2002; Clarkson \& Zunckel 2010; Huterer \& Shafer 2017) for a review. 
We approach this problem by attempting to construct potential for a
given redshift dependence of the equation of state parameter $w(z)$
for the dark energy component.
We do this for both quintessence and tachyon models: while a number of
solutions exist for quintessence models (Scherrrer 2015; Battye \& Pace 2016), few solutions are available
for tachyon models. In Scherrer (2015), a mapping between CPL parameters and potentials is explored while an analytic approximation for various scalar field models is obtained by Battye \& Pace (2016).

In \S{\ref{sec:eqns}}, we set up equations for tachyon and
quintessence models.
In \S{\ref{subsec:cwt}} and \S{\ref{subsec:cwq}}, we do reconstruction
of potential for $w(z) = constant$.
In \S{\ref{sec:gen_w}}, we outline the numerical recipe for reconstruction
for any general $w(z)$ and illustrate it with results for some simple
cases.  

\section{Basic Equations}
\label{sec:eqns}

We are interested in late time evolution of the Universe.
Given observations that indicate that the spatial curvature is
consistent with zero, and that radiation does not contribute to the
expansion history at $z \leq 100$, we choose to work with only matter
and dark energy.
The method we outline can be generalized without any modifications to
include other cases.  
For illustration of the method, we work with the CPL
parameterization (Chevallier \& Polarski 2000; Linder 2002).
The functional form for $w(z)$ is defined in terms of two constants,
which we call $p$ and $q$: 
\begin{equation}
    w=p+q(a-a_i)  \label{eq1}
\end{equation}
$p$ is the value of $w$ at some $t=t_i$ while $q$ gives
rate of change of $w$ with scale factor.
Symbols $w_0$ (for $p$) and $w_1$ (for $q$) are often used while using
this parameterization, if $t_i$ is taken to be the present time $t_0$. 
Continuity equation for dark energy density ${{\rho}}_{de}$ is:
\begin{equation}
    \frac{d{\rho}_{de}}{dt} = -3(1+p+q(a-a_i))\frac{\dot
      a}{a}{\rho}_{de}   \label{eq2}
\end{equation}
Using this equation, we get:
\begin{equation}
    {\rho}_{de} = {\rho}_{de}^i
    \left(\frac{a_i}{a}\right)^{3(1+p-qa_i)} \exp[-3q(a-a_i)]
    \label{eq3}
\end{equation}
where ${\rho}_{de}^i$ is density at some initial time. 
From now on we use a scaled dimensionless variable for time:
$t=tH_i$.
Friedmann equation then takes the form:
\begin{equation}
    \frac{\dot a^2}{a^2} = \frac{\alpha}{a^3} +
    \frac{\beta}{a^{3(1+p-qa_i)}e^{3qa}} 
    \label{eq4}
\end{equation}
where $\alpha$ and $\beta$ are constants defined as:
\begin{equation}
  \alpha = \Omega_{m\, i}\quad \beta = (1-\Omega_{m\,
    i})a_i^{3(1+p-qa_i)}e^{3qa_i} 
  \label{eq5}
\end{equation}
These are related to the density parameter for  matter and dark energy
at the initial time.

\subsection{Tachyon field}

Tachyon models for dark energy have an action of the following form: 
\begin{equation}
I = \int{d^4x\sqrt{-g}\left[-V(\phi)\sqrt{1-\partial^\mu\phi \partial_\mu\phi}\right] } 
\end{equation}

In these models the energy density and pressure can be written as: 
\begin{eqnarray}
\rho_\phi &=&
              \frac{V(\phi)}{\sqrt{1-\partial^\mu\phi \partial_\mu\phi}}
              \nonumber \\
P_\phi &=& -V(\phi)\sqrt{1-\partial^\mu\phi \partial_\mu\phi}
           \nonumber 
\end{eqnarray}
For these models the equation of state parameter is related to the
time derivative of the field as $w = -1 + {\dot{\phi}}^2$ for a
homogeneous field.
Thus we have:
\begin{equation}
    \frac{d\phi}{dt} = \sqrt{1+p+q(a-a_i)}
    \label{eq6}
\end{equation}
Combining eq.\eqref{eq4} and eq.\eqref{eq6} 
\begin{equation}
  \phi(a) = \int \frac{\sqrt{a(1+p+q(a-a_i))}}{\sqrt{\alpha +
      \frac{\beta}{a^{3p-qa_i}e^{3qa}}}}da 
  \label{eq7}
\end{equation}
Using the relation between the energy density and the potential, we
can write:
\begin{equation}
  V(\phi) = \sqrt{-w}\rho_{de}
  \label{eq8}
\end{equation}
Since we know $\rho_{de}$ as a function of $a$ from eq.\eqref{eq3}, we
can compute $V(a)$.
The combination of Eqn.\ref{eq7} and Eqn.\ref{eq8} gives a parametric
solution for the potential as a function of the field $\phi$, with the
scale factor $a$ playing the role of the intermediate parameter. 

\subsection{Tachyon field: Constant $w$}
\label{subsec:cwt}

We start by considering the special case of $w=$~constant, i.e.,
$q=0$. 
The integral in equation \eqref{eq7} takes following form for constant $w$:
\begin{equation}
    \phi(a) = \int \frac{\sqrt{a(1+w)}}{\sqrt{\alpha + \frac{\beta}{a^{3w}}}}da
    \label{eq9}
\end{equation}
Defining:
\begin{equation}
    x^2 = \alpha+\frac{\beta}{a^{3w}}
    \label{eq10}
\end{equation}
reduces the integral to form:
\begin{equation}
    \phi(x) = \int\frac{\sigma}{(x^2-\alpha)^k}dx 
    \label{eq11}
\end{equation}
where $\sigma$ and $k$ are:
\begin{equation}
    \sigma = -\frac{2\sqrt{1+w}}{3w\beta}\beta^k,\quad \quad k = \frac{w+\frac{1}{2}}{w}
    \label{eq12}
\end{equation}
Integral in eq.\eqref{eq11} is trivial for $w=-\frac{1}{2}$ where we get:
\begin{equation}
    \phi(a) = \sigma\sqrt{\alpha + \beta a^{3/2}}
    \label{eq13}
\end{equation}
Potential $V(a)$ for constant $w$ case is:
\begin{equation}
    \frac{V(a)}{H_i^2} = \frac{3\sqrt{-w}\beta}{8\pi G a^{3(1+w)}}
    \label{eq14}
\end{equation}
When $w=-\frac{1}{2}$, we get:
\begin{equation}
  \frac{V(\phi)}{H_i^2} =    \frac{3\beta}{8\pi G\sqrt{2}\left[
      \frac{\phi^2}{\beta \sigma^2}-\frac{\alpha}{\beta} \right]} 
    \label{eq15}
\end{equation}
For other values of $w$, integral in equation \eqref{eq9} does not
have a closed form solution.
The result can be expressed in the form of hypergeometric functions:
\begin{equation}
  \begin{split}
    \phi(a) =& \frac{2 a}{3} \left[ \frac{{a \left(1+w\right)}\left({\beta  a^{-3
            w}+\alpha }\right)}{\alpha \left(\beta  a^{-3 w}+\alpha
      \right)} \right]^{1/2} \\
  & \times \, \, _2F_1\left[\frac{1}{2},-\frac{1}{2
    w};1-\frac{1}{2 w};-\frac{a^{-3 w} \beta }{\alpha }\right]
\label{eq16}
\end{split}
\end{equation}
From eq.\eqref{eq14}, we have $V(a)$, we need to invert
eq.\eqref{eq16} to get $a(\phi)$ and substitute it in equation
\eqref{eq14} to get $V(\phi)$.
Please note that for background calculations one does not really need
$V(\phi)$, $V(a)$ contains the relevant information.
However for a study of spatial perturbations we require $V(\phi)$ as
$\phi$ can take on different values at different points at a given
time. 
A number of numerical libraries provide routines for calculation of
$_2F_1(a,b,c,g)$.
GNU Scientific library has function {\tt gsl\_sf\_hyperg\_2F1}, which
computes $_2F_1(a,b,c,g)$ for $|g|<1$.
In case of eq.\eqref{eq16}, $g<0$ and for extending to $g<-1$, there
are standard transformations available in literature(see
Pearson's thesis (2009) for a detailed account of computation of
hypergeometric functions, we use transformations mentioned in section
4.6 of Pearson's thesis (2009)).
For $g=-\frac{a^{-3 w} \beta}{\alpha }<-1$, we use following formulae
for computing $_2F_1(a,b,c,g)$:  
\begin{equation}
\small
\begin{split}
_2F_1(a,b,c,g) = & \frac{1}{(1-g)^a}
\frac{\Gamma(c)\Gamma(b-a)}{\Gamma(b)\Gamma(c-a)}
\\&_2F_1(a,c-b,a-b+1,\frac{1}{(1-g)})\\ 
&+\frac{1}{(1-g)^b} \frac{\Gamma(c)\Gamma(a-b)}{\Gamma(a)\Gamma(c-b)}
\\&_2F_1(b,c-a,b-a+1,\frac{1}{(1-g)}) 
\end{split}
\label{eq17}
\end{equation}
Equation \eqref{eq9} can be written in the form of a differential
equation which makes its relationship with other functions clear. 
Let
\begin{equation}
   g=  -\frac{a^{-3 w} \beta
   }{\alpha }
   \label{eq18}
\end{equation}
Then eq.\eqref{eq9} can be differentiated to obtain:
\begin{equation}
    g(1-g)\frac{d^2\phi}{dg^2} + \left[\left(\frac{1}{2w}+1\right)
      -\left( \frac{3}{2}+\frac{1}{2w}\right)g
    \right]\frac{d\phi}{dg}=0 
    \label{eq19}
\end{equation}
It can be integrated twice to obtain $\phi(g)$ in terms of incomplete
beta functions $B(g;a,b)$, which are related to $_2F_1(a,b,c,g)$: 
\begin{equation}
\phi(g) = C_1B(g;1-u,1+u+v) + C_2
\label{eq20}
\end{equation}
where $C_1$, $C_2$ are constants of integration and 
\begin{equation}
    \begin{split}
        u =& \left(\frac{1}{2w}+1\right)\\
        v =&-\left( \frac{3}{2}+\frac{1}{2w}\right)\\
    \end{split}
    \label{eq21}
\end{equation}
$B(g;a,b)$ is related to $_2F_1(a,b,c,g)$ (Weisstein webpage 2018)
as follows:
\begin{equation}
    B(g;a,b) = \frac{g^a}{a}{_2F_1}(a,1-b,a+1,g)
    \label{eq22}
\end{equation}
We can invert either eq.\eqref{eq16} or eq.\eqref{eq22} to obtain
$a(\phi)$ and then us eq.\eqref{eq14} to obtain $V(\phi)$.
We have used the Newton-Raphson method for inversion from $\phi(a)$ to
$a(\phi)$ and then on to $V(\phi)$.
This is useful in dynamically calculating $V(\phi)$ and the derivative 
$V_{\phi}(\phi)$ when $\phi$ has spatial variations in presence of
perturbations. 

\begin{figure}
    \centering
    \includegraphics[width=.8\columnwidth]{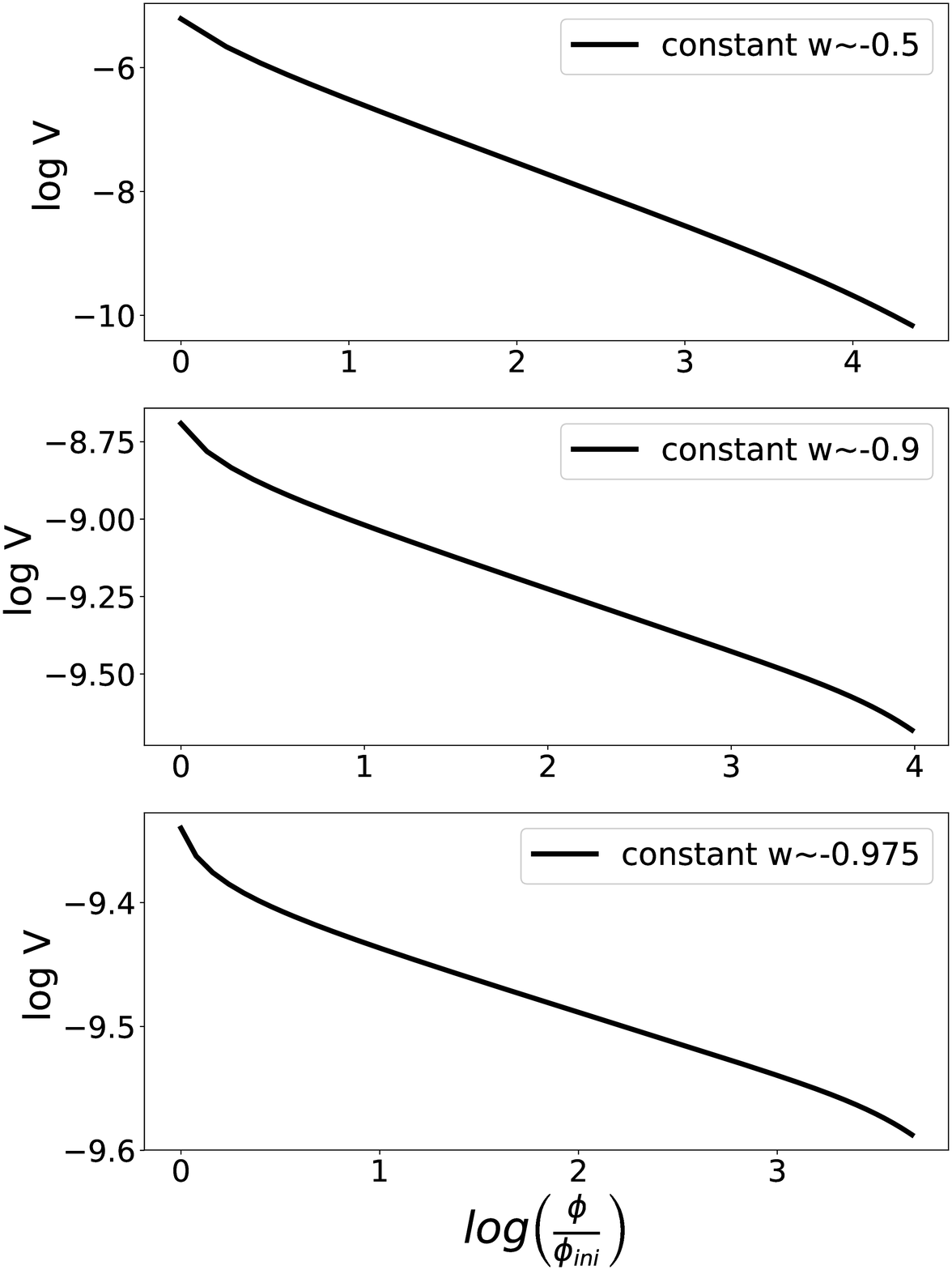}
    \caption{We plot tachyon potentials simulated,
      for constant $w$, using methods described in previous
      section. Different panels are for different constant w values.}  
    \label{fig:fig1}
\end{figure}

\begin{figure}
    \centering
    \includegraphics[width=.9\columnwidth]{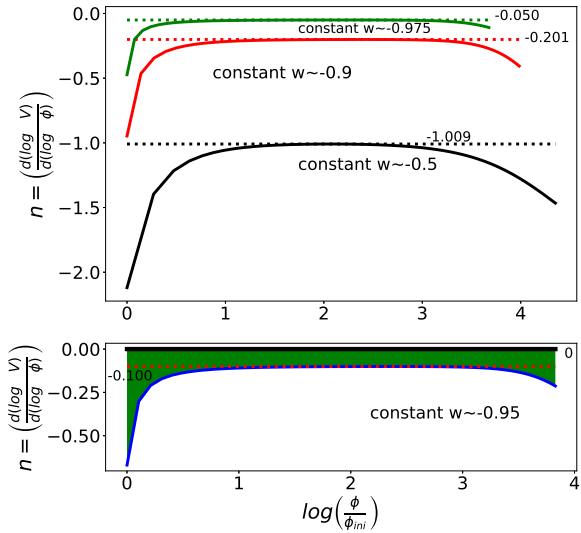}
    \caption{We plot the slope of the potential as a function of the
      field.
      From this log-log plot we can see that there is a almost flat
      plateau with deviations at two ends.
      Thus the potential is close to a power law. In upper panel we plot three cases of constant $w$. Lower panel is for 3-$\sigma$ constrained boundaries(-1.0,-0.95) as described in text. The shaded region is allowed set of potentials as per the constraints found in Tripathi
    {\em et al.} (2017).
      The value of constant central part changes with value of
      constant $w$.}  
    \label{fig:fig1b}
\end{figure}

\subsubsection{Form of the potential for constant $w$}

Here we plot(figure \ref{fig:fig1}) the potential $V(\phi)$ for
different values of $w$. 
We can see from the plot that the dependence of $V(\phi)$ is close to
a power law. To get insight into this behaviour, we plot derivatives of
log of potential with respect to log of field in figure
\ref{fig:fig1b}. We see that in central part there is approximate flat
curve indicating that in this region the potential can be approximated
by power laws.

\begin{figure}
    \centering
    \includegraphics[width=.8\columnwidth]{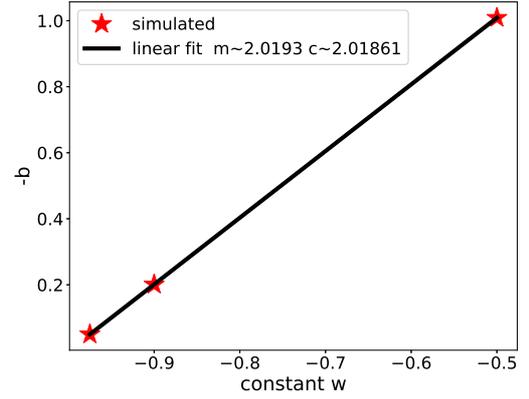}
    \caption{For different $w=constant$ values, we obtain the approximate b for central linear part(as marked in figure\ref{fig:fig1b}). As shown here, b values follow
      a linear relation with $w$. the fitted line has slope $m=2.3163$
      and intercept $c=2.30258$.} 
    \label{fig:fig2}
\end{figure}

We can approximate potential in this flat region with form:
\begin{equation}
    V(\phi) = c\phi^b \label{eqf1}
\end{equation}
For this form we have done fitting for different values of constant
$w$ and then we find the relationship between constant $w$ and $b$ which is
linear as shown in \ref{fig:fig2} . These fittings are crude given that evolution of $w$ and other
quantities is very sensitive to form of potential.  

     {This can
    potentially be used to constrain the potential for tachyon fields if
    one already has observational constraints on $w$.
    We are working on a detailed analysis of observational constraints
    to be presented in a forthcoming publication, here we present an
    example of such an exercise. We make use of existing studies of
    observational constrains on $wCDM$  models. In one such study,
    Tripathi {\em et al.} (2017) 
    combined the results from 3 different data sets to obtain $3\sigma$
    confidence intervals for constant $w$ and CPL $w(z)$ models. We
    use confidence intervals for the constant $w$ while working in the
    regime $w \geq -1$, i.e., we use the confidence interval
    ($-1.0,-0.95$) and reconstruct corresponding potential slope in
    lower panel of \ref{fig:fig1b}.  This is shown as a shaded region
    and marks the allowed slope for the potential.  This is a
    simplistic approach and we are working on a detailed analysis
    while accounting for possible variations of other cosmological
    parameters.} 

\subsection{Quintessence}
\label{subsec:cwq}

The action for quintessence field is:
\begin{equation}
I = \int{d^4x\sqrt{-g}\left[\frac{1}{2}g^{\mu \nu} \partial_{\mu} \phi \partial_{\nu} \phi \,-\, V(\phi)\right] }
\end{equation} 
with effective pressure and density:
\begin{eqnarray}
\rho_\phi = \frac{\dot{\phi}^2}{2} + V(\phi)\\
P_\phi = \frac{\dot{\phi}^2}{2} - V(\phi)\\
w_\phi = \frac{P_\phi}{\rho_\phi}=\frac{\dot{\phi}^2 - 2V}{\dot{\phi}^2 + 2V}
\end{eqnarray}

For Quintessence models of dark energy, $w$ is related to time
derivative of the  field and the potential, and we have: 
\begin{equation}
  \frac{d\phi}{dt} =  \sqrt{(1+w)\rho_\phi} =\sqrt{(1+p+q(a-a_i))\rho_\phi}
  \label{eq23}
\end{equation}
where  
\begin{equation}
  V(\phi) = \frac{1}{2}(1-w)\rho_\phi=\frac{1}{2}(1-p-q(a-a_i))\rho_\phi
  \label{eq24}
\end{equation}
From equations \eqref{eq3} and \eqref{eq23}, we obtain:
\begin{equation}
  \frac{d\phi}{dt} = \sqrt{(1+p+q(a-a_i))\frac{3}{8\pi
      G}\frac{\beta}{a^{3(1+p-qa_i)}}e^{-3q(a-a_i)}} 
  \label{eq25}
\end{equation}
Equation \eqref{eq25} can be combined with eq.\eqref{eq4} to obtain
$\frac{d\phi}{da}$.
For potential we have from \eqref{eq24} and \eqref{eq3}:
\begin{equation}
  \frac{V}{H_i^2} = \frac{3}{2}\frac{(1-w)}{8\pi
    G}\frac{\beta}{a^{3(1+p-qa_i)}}e^{-3q(a-a_i)} 
  \label{eq26}
\end{equation}
This system of equations specifies the solution.

\subsection{Quintessence field: Constant w}

For $w(a)=constant$, we obtain a closed formula for $V(\phi)$(see Sangwan {\em et al.} (2018) and references within for previous work on this).
In this case, eq.\eqref{eq25} reduces to:
\begin{equation}
  \frac{d\phi}{dt} = \sqrt{(1+w)\frac{3}{8\pi G}\frac{\beta}{a^{3(1+w)}}}
  \label{eq27}
\end{equation}
and
\begin{equation}
  \frac{d\phi}{da} =\sqrt{\frac{3(1+w)}{8\pi G}} 
  \sqrt{\left[ \frac{1}{\frac{\alpha a^{3w}}{\beta}+1}\right]}\left(
    \frac{1}{a}\right) 
  \label{eq28}
\end{equation}
Defining:
\begin{equation}
    \lambda = \sqrt{\frac{3(1+w)}{8\pi G}} 
    \label{eq29}
\end{equation}
and
\begin{equation}
     x^2 = \frac{\alpha a^{3w}}{\beta}+1
    \label{eq30}
\end{equation}
We have, 
\begin{equation}
    \phi(x) = C_1 + \frac{2\lambda}{3w}\int\frac{dx}{x^2-1}  
    \label{eq31}
\end{equation}
here $C_1$ is a constant of integration.
The solution is :
\begin{equation}
    \phi(x) = -\frac{\lambda}{3w}[\log(1+x) - \log(x-1)]
    \label{eq32}
\end{equation}
Inverting this we get:
\begin{equation}
    x = \frac{e^{-3w\phi/\lambda}+1}{e^{-3w\phi/\lambda}-1}
    \label{eq33}
\end{equation}
Defining:
\begin{equation}
    m = -\frac{3w\phi}{2\lambda}
    \label{eq34}
\end{equation}
We rewrite eq.\eqref{eq33}:
\begin{equation}
    x = \coth{m}
    \label{eq35}
\end{equation}
And we get
\begin{equation}
    a^{3w} = \frac{\beta}{\alpha}\left[(\coth{m})^2 -1\right]
    \label{eq36}
\end{equation}
Substituting this in eq.\eqref{eq26},
\begin{equation}
    \frac{V(\phi)}{H_i^2} = \frac{3(1-w)\beta}{16\pi G}
    \left[ \frac{\beta}{\alpha}((\coth{m})^2 -1) \right]^{-\frac{(1+w)}{w}}
    \label{eq37}
\end{equation}
Equivalently,
\begin{equation}
    \frac{V(\phi)}{H_i^2} = \frac{3(1-w)\beta}{16\pi G}
    \left[ \frac{\alpha}{\beta}
    \sinh^2{\left(-\frac{3w\phi\sqrt{8\pi G}}{2\sqrt{3(1+w)}}\right) }
     \right]^{\frac{(1+w)}{w}}
    \label{eq38}
\end{equation}
Derivations for constants $w$ case for quintessence and phantom models
were done by Sangwan {\em et al.}(2018) and they obtain the same form
for quintessence models as in
eq.\eqref{eq38}. {The reconstruction approach can be
    used to 
    constrain potentials from observations, as was done in Sangwan
    {\em et al.}(2018). They used the constrained ranges from Tripathi
    {\em et al.} (2017), to constrain the quintessence potentials for
    constant $w$. They also constrain the potentials for CPL and
    logarithmic $w(z)$ using some approximations. Their work can be
    numerically generalized, using formalism developed in this
    article, to various $w(z)$ for both tachyonic as well as
    quintessence models.} 

\section{General case}
\label{sec:gen_w}

For an arbitrary function $w(a)$, continuity equation for that
component is: 
\begin{equation}
    \frac{d\rho_\phi}{\rho} = -\frac{3(1+w)}{a}da 
    \label{eq39}
\end{equation}
giving
\begin{equation}
    \rho_{\phi} = \rho_{{\phi}_i}
    \exp\left[-3\int\frac{1+w}{a}da\right] 
    \label{eq40}
\end{equation}
Equivalently
\begin{equation}
    \Omega_\phi := \frac{8\pi G\rho_{\phi}}{3H_i^2} =
    \Omega_{{\phi}_i} e^{-3\int\frac{1+w}{a}da} 
    \label{eq41}
\end{equation}
Subscript i represent values at some initial time.

\begin{figure}
  \centering
  \includegraphics[width=1\columnwidth]{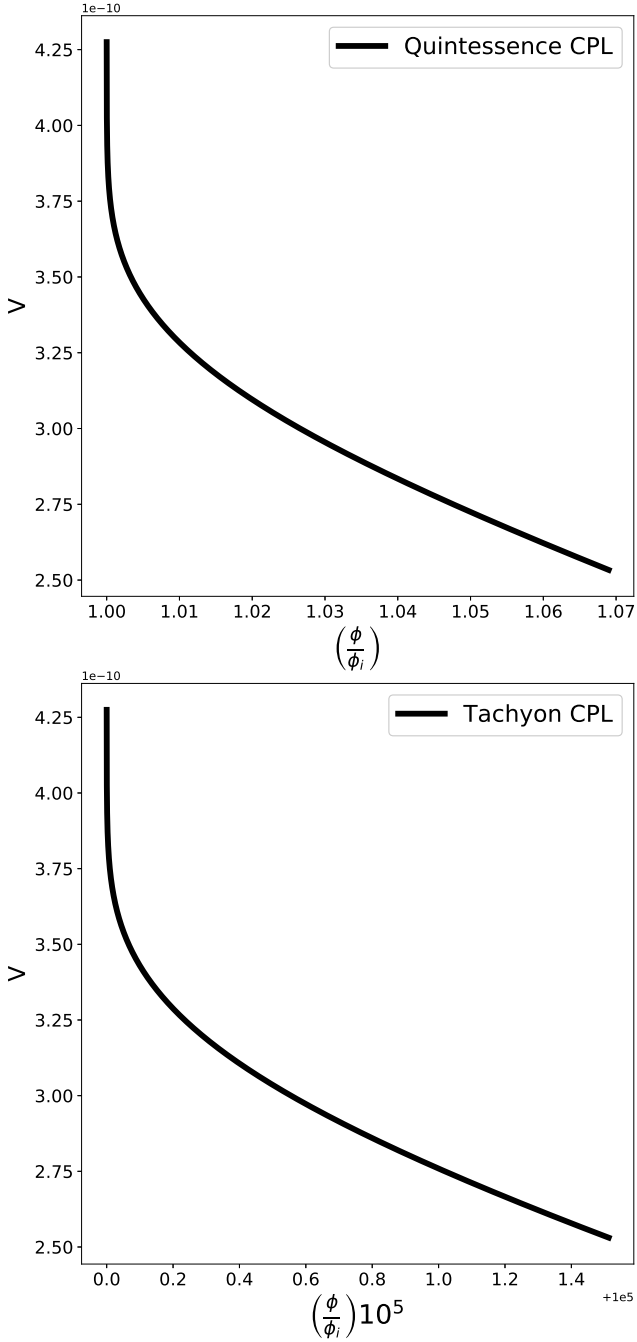}
  \caption{$V(\phi)$ simulated for CPL parameterization for
    quintessence and tachyon models. The shape of curve is same for
    quintessence and tachyon field but the rate of evolution of
    field very different. Field traverses longer distances in field
    space for quintessence case. This might have interesting
    implications in context of Swampland criteria of String theory.} 
  \label{fig:fig5}
\end{figure}

Using this evolution equation for energy density we can write
differential equations for tachyon and quintessence fields: 
\begin{equation}
  \frac{d\phi_{tach}}{da} = \frac{\sqrt{1+w}}{\sqrt{\frac{\alpha}{a} +
      a^2 \Omega_{\phi tach} }} 
  \label{eq42}
\end{equation}
\begin{equation}
  \frac{d\phi_{q}}{da} = \frac{\sqrt{3(1+w)\Omega_{\phi q}}}{\sqrt{8\pi
      G}\sqrt{\frac{\alpha}{a} + a^2 \Omega_{\phi q} }} 
  \label{eq43}
\end{equation}
where $\Omega_{\phi q}$  and  $\Omega_{\phi tach}$
are quintessence and  
tachyon field density parameters scaled as shown in eq.\eqref{eq41}
respectively. 
The potentials for two fields are:
\begin{equation}
  \frac{V(a)}{H_i^2} = \frac{3(1-w)\Omega_{\phi q}}{16\pi G}
  \label{eq44}
\end{equation}
\begin{equation}
  \frac{V(a)}{H_i^2} = \frac{3\sqrt{-w}\Omega_{\phi tach}}{8\pi G}
  \label{eq45}
\end{equation}
One can numerically integrate equations \eqref{eq41} and
\eqref{eq42}/\eqref{eq43} to get $\phi(a)$ and alongside use
\eqref{eq44}/\eqref{eq45} to obtain $V(a)$.
Hence one can obtain a numerical table of $V(\phi)$ vs $\phi$ in
desired range.
This table can be used for numerical fitting or interpolation
functions. 
For example, cubic splines(see book by Antia (2012)) can be used for fitting
to obtain spline coefficients which can be used for calculating
$V(\phi)$ and its gradients given a value of $\phi$.
Once we have spline coefficients and $\phi$, task is to find the
interval in which the value of $\phi$ lies so that we can use
coefficients corresponding to that interval.
Evaluation of the function can be time consuming, but the fact, that
for background values $\phi$ there is a correspondence between $\phi$
and $a$, comes to our rescue.
Typically perturbations have a small amplitude and hence deviation
from background in a particular simulation domain is small, and this
can be used to guess spline interval in that region.
For example, one might be simulating a spherical collapse in real
space and perturbations may be really strong towards centre but they
merge into background as one moves away from centre.
In this case for large radii, interval can be guessed from background
and then one can move toward smaller radii.
In this way for each new inner point one has to only search in the
adjacent intervals for interpolation if the field is continuous.
As an example we show here CPL potentials for quintessence and
tachyon field in \ref{fig:fig5}. The form obtained is similar to that obtained by Scherrer(2015).

\section{Coupled Quintessence mimicking $\Lambda CDM$}

Minimally coupled quintessence models(Amendola 1999; Shahalam {\em et al.} 2015) can exactly mimic $\Lambda CDM$
only with a completely flat potential, that is no field dynamics is
involved and equations just reduce to that in case of $\Lambda$.
However if energy exchange is allowed between quintessence field and
dark matter, a $\Lambda$ like evolution is possible even with field
dynamics and a time varying $w$. 
In this section we consider a quintessence model with following type
of coupling(Barros {\em et al.} 2018): 
\begin{equation}
  ^{\phi}T^{\mu}_{\nu},_{\mu} = Q\sqrt{8\pi G}\phi,_{\nu}\rho_{cdm}
  \label{eq46}
\end{equation}
\begin{equation}
  ^{c}T^{\mu}_{\nu},_{\mu} =-Q\sqrt{8\pi G}\phi,_{\nu}\rho_{cdm}
  \label{eq47}
\end{equation}
Please note a bit different notation in this section as described below.
$Q$ is the coupling constant between matter and Quintessence.
Subscript $lcdm$ denotes quantity corresponding to $\Lambda CDM$ and
$cdm$ subscript is for corresponding quantities for cold dark matter in
model with field, e.g. $\rho_{cdm}$ is density for cold dark matter in
model with an interacting dark energy field while $\rho_{lcdm}$ is
cold dark matter density as evolved within $\Lambda CDM$. Also $\Omega_c^i$ is density parameter for dark matter at initial time and $\Omega_\Lambda^i$ is $\Lambda$ counterpart.
Basic equations for this type of coupled model mimicking $\Lambda CDM$
were derived in Barros {\em et al.} (2018).
They write the potential $V(\phi)$ in terms of other variables, and do
not specify exact formula for $V(\phi)$.
Here we start from the equations derived in Barros {\em et al.} (2018)
and then reconstruct the formula for potential that gives the required
$\Lambda$ like behaviour.
For a field model giving same $a(t)$ as that of $\Lambda CDM$, we have:
\begin{equation}
  \left(\frac{\dot{a}}{a}\right) = \left(\frac{\dot{a}}{a}\right)_{\Lambda CDM}
  \label{eq48}
\end{equation}
Ignoring baryons and radiation we have:
\begin{equation}
  \rho_{cdm} +\rho_\phi = \rho_{lcdm} +\rho_\Lambda
  \label{eq49}
\end{equation}
and 
\begin{equation}
  p_\phi = p_\Lambda = -\rho_\Lambda
  \label{eq50}
\end{equation}
Combining the two, we have:
\begin{equation}
  \dot{\phi}^2 = \rho_{lcdm}-\rho_{cdm}
  \label{eq51}
\end{equation}
Continuity equation for matter is:
\begin{equation}
  \dot{\rho_{cdm}} + 3H\rho_{cdm} = -Q\sqrt{8\pi G}\dot{\phi}\rho_{cdm} 
  \label{eq52}
\end{equation}
giving:
\begin{equation}
  \rho_{cdm} = \rho_{cdm}^i\frac{a_i^3}{a^3}e^{-Q\sqrt{8\pi G}\phi}
  \label{eq53}
\end{equation}
Using \eqref{eq51} and \eqref{eq53} along with standard Friedmann
equation for $\Lambda CDM$, we get: 
\begin{equation}
  \frac{d\phi}{da}= \sqrt{\left( \frac{3}{8\pi G}
    \right)}\left(\frac{1}{a}\right)\frac{\sqrt{ 1-e^{-Q\sqrt{8\pi
          G}\phi} }}{\sqrt{1+\frac{\Omega_\Lambda^i a^3}{\Omega^i_c
        a^3_i}}} 
  \label{eq54}
\end{equation}
Arranging and integrating equation we obtain:
\begin{equation}
  \begin{split}
  \omega\quad & \log\left[\sqrt{e^{Q\sqrt{8\pi G}\phi}-1} + e^{Q\sqrt{8\pi G}\phi/2} \right]=\\
   & \log\left[\frac{\sqrt{1+\frac{\Omega_\Lambda^i a^3}{\Omega^i_c
        a^3_i}} -1}{\sqrt{1+\frac{\Omega_\Lambda^i a^3}{\Omega^i_c
        a^3_i}} + 1} \right]
    \end{split}
  \label{eq55}
\end{equation}
where $\omega = \pm\frac{2\sqrt{3}}{Q}$(- for negative $Q$)\\
Writing $a(\phi)$ as a function of $\phi$:
\begin{equation}
  a^3(\phi) = \frac{4C_1\Omega_c^i a^3_i}{\Omega_\Lambda^i}
  \left[ \frac{ f(\phi)^\omega}{(1-C_1 f(\phi)^\omega)^2} \right] 
  \label{eq56}
\end{equation}
with $C_1$ taking care of any constant of integration and
\begin{equation}
f(\phi) = \sqrt{e^{Q\sqrt{8\pi G}\phi}-1} + e^{Q\sqrt{8\pi G}\phi/2}
 \label{eq56b}
\end{equation}

\begin{figure}
    \centering
    \includegraphics[width=1\columnwidth]{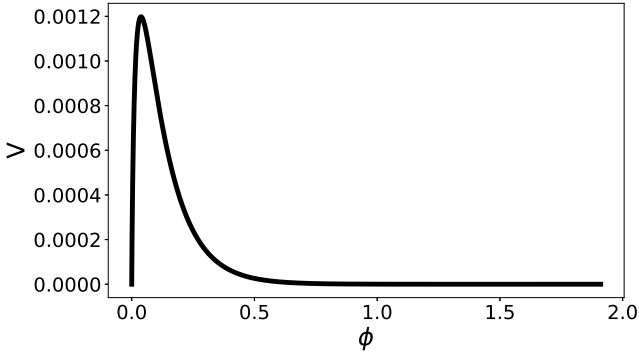}
    \caption{$V(\phi)$ for coupled quintessence mimicking $\Lambda
      CDM$ in background kinematics.} 
    \label{fig:fig6}
\end{figure}

Potential $V$ can be obtained from equations \eqref{eq50},
\eqref{eq51} and \eqref{eq53}: 
\begin{equation}
  \frac{V(\phi)}{H_i^2} = \frac{3\Omega_c^i}{8\pi G}
  \left[\frac{a_i^3}{2(a^3(\phi))}(1-e^{-Q\sqrt{8\pi G}\phi}) +
    \frac{\Omega_\Lambda^i}{\Omega^i_c}\right] 
  \label{eq57}
\end{equation}
Where $a^3(\phi)$ has a functional form as mentioned in \eqref{eq56}.
Form of potential is illustrated in \ref{fig:fig6}.
\begin{equation}
 \begin{split}
 & V,_\phi =\frac{3\Omega_c^i}{8\pi G} \\& \left[ \frac{Q\sqrt{8\pi
        G}e^{-Q\sqrt{8\pi G}\phi}a_i^3}{2a^3} -
    \frac{a_i^3}{2a^6}\frac{d(a^3)}{df}\frac{df}{d\phi}(1-e^{-Q\sqrt{8\pi
        G}\phi}) \right] 
        \end{split}
\end{equation}

Studies of perturbations in the coupled quintessence models can play
an important role in distinguishing these from $\Lambda$CDM models.
Our analysis of such models will be reported elsewhere. 

\section{Summary}
\label{sec:summary}

In this work we have described basic equations for reconstructing
potentials for quintessence and tachyon field.
We have given results for $w=constant$ case.
We show that analytical closed formulas are possible for quintessence
potentials in these cases while for tachyon fields such formulae are
obtained only for $w=-0.5$ case.
For other values of constant $w$, we provide formulae for numerical
reconstruction.
We also find a rough approximation to these constant $w$ potentials
for tachyon dark energy.
We describe numerical methods for numerical construction of tachyon
and quintessence potentials for arbitrary $w(a)$.
From numerical calculation of potentials for CPL cases for
quintessence and tachyon we show that the shape pf potential is same
for both of these, but the field rolls much more in quintessence case
than in tachyon case.
This could motivate further investigations in context of String
Swampland (Heisenberg {\em et al.} 2018; Agrawal {\em et al.} 2018; Akrami {\em et al.} 2018).
We have also studied coupled quintessence models.

The results of this study can be used for analysis of perturbations in
such models.
In particular we can compare growth of perturbations in models of
different types that have the same expansion history. 
We will report results on spherical collapse in perturbed dark
energy models in a forthcoming publication.

\section*{Acknowledgements}

Authors thank Dr. Ankan Mukherjee, Dr. H. K. Jassal, Dr. Varadharaj
Srinivasan and Avinash Singh for useful discussions.
This research has made use of NASA's Astrophysics Data System
Bibliographic Services. 

\vspace{-1em}



\end{document}